\documentclass[12pt]{article}
\usepackage[usenames,dvipsnames,svgnames,table,x11names]{xcolor}
\usepackage{kpfonts}
\usepackage{packages-csm}
\usepackage{simeontex-csm}
\usepackage{definitions-csm}
\usepackage[utf8]{inputenc} 
\usepackage{tikz}
\usetikzlibrary{arrows,decorations.markings}
\usetikzlibrary{calc}
\usepackage{slashed}
\usepackage{braket}
\usepackage[affil-it]{authblk}
\usepackage[outdir=../fig/]{epstopdf}
\usepackage[english]{babel}

\definecolor{refkey}{rgb}{0.85, 0.0, 0.3}
\definecolor{labelkey}{rgb}{0.2,0.2,0.6}

\def\prsqd{^{\prime 2}}

\newenvironment{PurpleEnv}%
{\color{Purple}}%
{\color{Black}}
\def\bpe{\begin{PurpleEnv}}
\def\epe{\end{PurpleEnv}}
{\color{Red}}%
{\color{Black}}

\def\ok{}

\preprint{{\tt IPMU18-0069}}

\title{Observables in Inhomogeneous Ground States at Large Global Charge}

\author[1]{Simeon Hellerman}
\author[1,2]{Nozomu Kobayashi}
\author[1,2]{Shunsuke Maeda}
\author[1,2]{Masataka Watanabe}
\affil[1]{\small Kavli Institute for the Physics and Mathematics of the Universe (WPI),The University of Tokyo Institutes for Advanced Study, The University of Tokyo, Kashiwa, Chiba 277-8583, Japan}
\affil[2]{\small Department of Physics, Faculty of Science, The University of Tokyo, Bunkyo-ku, Tokyo 133-0022, Japan}
\date{}

\allowdisplaybreaks[4]

\def\llok{\llsk\ok}

\begin{document}

\hypersetup{linkcolor=black}
  \maketitle 
     \abstract{
As a sequel to previous work, we extend the study of the ground state configuration of the $D=3$, Wilson-Fisher conformal $O(4)$ model.
In this work, we prove that for generic ratios of two charge densities, $\rho_1/\rho_2$, the ground-state configuration is inhomogeneous and that the inhomogeneity expresses itself towards longer spatial periods.
This is the direct extension of the similar statements we previously made for $\rho_1/\rho_2\muchlessthan 1$.
We also compute, at fixed set of charges, $\rho_1,\, \rho_2$,  the ground state energy and the two-point function(s) associated with this inhomogeneous configuration on the torus.
The ground state energy was found to scale $(\rho_1+\rho_2)^{3/2}$, as dictated by dimensional
analysis and similarly to the case of the $O(2)$ model.
 Unlike the case of the $O(2)$ model, the ground also strongly violates cluster decomposition in the large-volume, fixed-density limit, with a two-point function that
is negative definite at antipodal points of the torus at leading order at large charge. 
}
      \newpage
\setcounter{tocdepth}{3}
\setcounter{secnumdepth}{4}
\tableofcontents
\hypersetup{linkcolor=PaleGreen4}
\newpage

\section{Introduction}

Global symmetries can give conformal field theories interesting and useful simplifications. 
Asymptotic expansions for dimensions and OPE coefficients of charged local operators in terms of large global charge are sometimes possible \cite{Hellerman:2015nra,Alvarez-Gaume:2016vff,Monin:2016jmo,Loukas:2016ckj}, to any given order, using a small number of unknown coefficients which come from terms in the effective Lagrangian in the large charge sector.
These are strikingly parallel to the large-spin expansion using the light-cone bootstrap\footnote{See also more recent work \cite{Alday:2016njk,Alday:2016jfr}.}
\cite{Komargodski:2012ek, Fitzpatrick:2012yx}, in spite of the fact that the method of large global charge includes the use of Lagrangian methods as opposed to the abstract conformal bootstrap. \ok\ok
These methods, of course, work best in the regime of large charge, and complement the regime of $O(1)$ charges and operator dimensions \cite{Kos:2016ysd, Kos:2015mba, Kos:2013tga, Dey:2016zbg, Diab:2016spb}, which is effectively accessed using the method of linear programming \cite{Rattazzi:2008pe , ElShowk:2012ht , El-Showk:2014dwa} to solve the bootstrap
equations. \ok\ok

In
\cite{Hellerman:2015nra,Alvarez-Gaume:2016vff,Monin:2016jmo},
the operator dimensions of charged local operators are calculated by
radially quantizing the large-charge effective Lagrangian on a spherical
spatial slice.  \ok\ok
The idea of these papers are as follows:
If we consider a system with large global charge $J$ and charge density $\rho$, on a spatial slice of the scale $R_{\rm geometry}$,
the large-charge effective Lagrangian has its UV scale at
$E\ll{\rm UV} \equiv \r\uu{{1\over{D-1}}}$ and  IR at $E\ll{\rm IR}=1/R_{\rm geometry}$,
and this large hierarchy, $E\ll{\rm IR} / E\ll{\rm UV} \propto J\uu{-{1\over{D-1}}}$,
renders the theory weakly coupled.
In other words, we can compute various physical quantities perturbatively
in terms of $1/J$, with all the quantum corrections and higher-derivative terms suppressed. \ok\ok

Before studying the operator dimensions and various physical quantities, the first thing necessary here is to know the structure of the large-charge effective Lagrangian and the nature of the ground state with a fixed set of charges.\ok\ok
In the limit where the charge goes to infinity $J\to\infty$, one may also take the size of the sphere to infinity ${\cal V} \to \infty$ and consider the system on a infinite flat space with fixed average charge density, $\rho \equiv J / {\cal V}$.  In this limit, many possibilities may be realized:
\begin{itemize}
\item[(a)]{
The ground state may become a homogeneous configuration, related directly to the thermodynamic limit in infinite volume,
at finite chemical potential and zero temperature. 
In some cases, this is indeed so, and various interesting new phases of matter
with spontaneously broken conformal and Lorentz symmetries have been derived, which describe such
breaking patterns \cite{Hellerman:2015nra,Alvarez-Gaume:2016vff,Monin:2016jmo}. 
}
\item[(b)]{The ground state may be homogeneous but the thermodynamic limit in infinite volume may not exist.  This possibility
is realized by some well-known and perfectly well-behaved CFT, including free complex scalar fields, and superconformal
theories with moduli spaces of supersymmetric vacua in flat space.  These CFT have discrete spectrum and perfectly well-defined thermodynamics on the sphere, but not in flat space; in the presence of curvature, the ground state at large $R$-charge
 is homogeneous and satisfies $T_{00} \propto ({\tt Ricci~scalar})\uu{+\hh}\cc \r$, and so the space of ground states collapses
 at zero curvature, even in finite volume.
 Examples include superconformal theories with vacuum manifolds, such 
 as the $3D$ $\mathcal{N}=2$ superconformal XYZ model at large $R$-charge and \cite{Hellerman:2017veg}, $4D$ $\mathcal{N}=2$ rank 1 SCFTs at large $R$-charge \cite{Hellerman:2017sur,Hellerman:2018xpi}.}
 \item[(c-IR)]{The ground state may be inhomogeneous in finite volume, with the scale of the inhomogeneity set by the scale of
 the geometry itself.  For instance, on a torus, the ground state may break the translational symmetry to $\IZ\ll {k\ll 1} \times
 \IZ\ll{k\ll 2}$ or $\IZ\ll k \times U(1)$, depending on the geometry of the torus and the details of the CFT.
 This case is partly studied in \cite{Hellerman:2017efx} and will be expanded in this paper later.
 }
 \item[(c-UV)]{The ground state may be inhomogeneous on the scale set by the charge density itself; this possibility would be realized, for instane, by a striped phase, with the periodicity set by $\r\uu{-\hh}$ times some fixed constant determined by the dynamics.  While we know
no examples of a relativistic CFT with such a ground state at large global charge, it is a logical possibility that may be realized in some not-yet-discovered theories.  (However see recent work \cite{Cuomo:2017vzg} in which the ground state with large charge
combined with large angular momentum can be shown to break translational invariance spontaneously at the scale of the charge
density itself, in certain limits.) }
 \end{itemize}
There is actually one other possibility that the ground state is not semiclassical at all, i.e., the fluctuations are not supressed by powers of $E_{\rm IR}/\mu$.
This by no means happens when all the degrees of freedom are coupled in a generic way.
You can nonetheless come up with several examples where this happens --
One is two completly decoupled CFTs, one of which the symemtry acts trivially, and another is just two-dimensional CFT with a global symmetry,
in which the current degress of freedom decouples because of Sugawara decomposition. 
Note that even in 2D CFTs with currents, when the theory has a continuous, rather than discrete, spectrum, the Sugawara decomposition doesn't apply and the theory does not realise this possibility.
To the contrary, this kind of theories generically shows a simplification at large charge limit, and has been studied in the context of relativistic strings in the Regge limit \cite{Gross:1987ar,Gross:1987kza,Polchinski:1991ax,Baker:1999xn, Sonnenschein:2018aqf,Hellerman:2017upi,Hellerman:2014cba}.
We will not further comment on this non-semiclassical possibility which is irrelevant to the topic of this paper.

Studying and classifying the above possibilities are quite important in checking the validity of the effective field theories at large global charge.
Especially, when the large-charge ground state realizes the case (c-UV), the suppression of quantum fluctuations in powers of $E_{\rm IR}/\mu$ is never there in the first place.
Reassuringly, though, there are several facts against (c-UV) both in the cases of $D=3$, $O(2)$ and $O(4)$ Wilson-fisher fixed points at large charge/set of charges.
We will collect evidences against (c-UV) in the first two parts of this paper.
In the first part, we are going to, prove that the $O(2)$ model and the $O(4)$ model with vanishing $\rho_1$ or $\rho_2$ (charge densities), at large charge realise possibility (a).
Also, we prove that for generic charge ratios in the $O(4)$ model on the torus, it is possible to eliminate the possibility that the ground state configuration is inhomogeneous in both cycles.

In the second part of this paper we will particularly concentrate on the Wilson-Fisher $O(4)$ fixed point in $D=3$ for any set of total charges, $J_{1,2} = \int \cc d\uu 2 \cc \r\ll {1,2}$, which are taken to be large in arbitrary fixed
ratio. (We partly answered the same question in the limit $J_1 / J\ll 2 \muchlessthan 1$ in \cite{Hellerman:2017efx}; in this regime the ground state is nearly homogeneous and
the inhomogeneity can be treated as a perturbation.)  This case is the simplest nontrivial example with an inhomogeneous ground state at
large global charge.  (Very interesting cases of inhomogeneous ground states combining large charge and angular momentum
have also recently been discovered \cite{Cuomo:2017vzg}.)
Extending the result of \cite{Alvarez-Gaume:2016vff},
we find that the ground state solution for large charges
in generic charge ratio is inhomogeneous with a family
of classical solutions periodic in one spatial direction; but, that the family of solutions has an energetic preference for longer spatial periods, which is eventually bounded by the longest scale of the geometry of the spatial slice.
This means that for any set of large charge densities the ground state configuration varies very slowly compared
to the scale of the charge density, so that the large-charge effective Lagrangian is parametrically reliable, and the possibility (c-IR) is internally consistent in this case.  Although this analysis supports the consistency of the possibility (c-IR) to be realised, it still does not rule out the possibility (c-UV) as we work in the regime of the EFT throughout.

In the last part of this note where the issue with inhomogeneity has been settled, we will compute the classical energy and the two-point functions associated with the ground state configuration on a torus spatial geometry, $\mathbb{R}_t\times T^2$ using the above result.
We show that the classical energy scales like $(\rho_1+\rho_2)^{3/2}$ as expected from previous studies, with subleading correction which goes as $(\rho_1+\rho_2)^{1/2}$.  For the ground state of the $O(2)$ model,
the $J\uu \hh$ term comes from the curvature coupling
on the sphere, and vanishes on the torus.  For the
$O(4)$ model, on the other hand,
there is a term scaling as $J\uu\hh$ is nonzero even in 
flat space: The $(\partial|\partial \chi|)\sqd / |\pp\chi|$ term,
which vanishes classically for the ground state solution
in the $O(2)$ model, does not vanish classically
for the inhomogeneous ground state solutions of the
$O(4)$ model with generic charges, and thus
makes a contribution of order $J\uu\hh$ even in the
absence of curvature.

{We also compute the two-point function on the torus from the inhomogeneous ground state, and see the resulting spatial dependence in term of operator insertions.
Interestingly, even at leading order, it exhibits a dramatic difference from that of the $O(2)$ model: The two-point function with insertions at two antipodal points is negative definite and nonzero at leading order in the charge, which can only occur
together with a breakdown of cluster
decomposition and spontaneous strong spatial inhomogeneity at the infrared scale.}
We hope these results can be checked against Monte-Carlo simulations as in \cite{Banerjee:2017fcx}.

\section{Goldstone counting and the (in)homogeneity of large-charge ground states}

As promised, we are going to derive several facts about (in)homogeneity of the ground state configuration at large charge. on the torus.
This can be done in a simple way by counting the number of Goldstone bosons and matching with the number of available light modes.

The following results can and should be generalised from torus spatial slice to the sphere spatial slice.
However, because we only deal with the torus time slice in this paper, we are not going to mention the sphere case.
Note that the sphere spatial slice adds a little more complexity to the problem because the symmetry group on it is not Abelian.
But this is not in any way an obstacle in expanding these results, as you can use the method of \cite{Watanabe:2013uya} even in the case of non-Abelian symmetries.

\heading{A comment on helical symmetries and chemical potentials}

Many discussions of finite-density ground states in the condensed matter literature, as well as some recent work on large quantum-number
expansions in CFT \cite{Alvarez-Gaume:2016vff},
make use of chemical potentials in order to describe the large charge ground state.  This is
natural in the thermodynamic limit though slightly less so in finite volume, where the legendre transform between
chemical potentials and densities must be replaced by a Fourier transform of the quantum partition function \cite{Loukas:2016ckj}.  

For
this and other reasons, in the present
work, we describe the ground state in terms of a classical solution with a helical symmetry, \it i.e., \rm a symmetry
under a combined time translation and global symmetry translation \cite{Hellerman:2015nra} \cite{Hellerman:2017veg}\cite{Baker:1999xn}\cite{Sonnenschein:2018aqf}.
 In classical mechanics the two notions are precisely
equivalent after a change of variables: The overall lowest-energy classical solution of the system with chemical potential is
always static on general grounds of Hamiltonian mechanics; therefore after a time-dependent global symmetry transformation
that removes the chemical potential term from the Hamiltonian, the lowest-energy ground state with a given charge must have
a helical symmetry.  Quantum corrections to the classical picture of the ground state can be added systematically in an
asymptotic expansion in the inverse charge \cite{Loukas:2016ckj}\cite{Hellerman:2017veg}.

The equivalence of these two descriptions of the ground state also emphasizes an important point that is sometimes ignored:
A chemical potential always preserves the same symmetries -- Lorentz and global symmetries -- of the system that
are preserved by the Hamiltonian without chemical potential: A constant chemical potential can be removed by a change
of variables.  The change in the ground state of the system induced by a chemical potential (at zero temperature) should
always be viewed as spontaneous rather than explicit breaking.  The description in terms of a helical solution merely emphasizes
this fact which is otherwise obscured by the nontrivial transformation law after the change of variables.

While the helical frequency is a spontaneous rather than explicit breaking,
the only light goldstone modes are those corresponding to symmetries commuting with the generator $\hat{g}$ defining the helical time-dependence $\exp{i \hat{g} \omega t}$,
since the helical ground state is only time-independent up to a symmetry transformation by $\hat{g}$.  The symmetries commuting with the helical frequency are precisely those which would commute with the chemical potential after the change of variables,
and it is only these that generate light goldstone modes.  The generators not commuting with the helical frequency are "massive goldstone bosons" \cite{Watanabe:2013uya} whose masses are above the cutoff but still precisely fixed by the symmetry algebra (See the supersymmetric version in the $W = \Phi^3 $ model for \cite{Hellerman:2015nra}, and the non-supersymmetric, $O(2N)$ version  \cite{Alvarez-Gaume:2016vff} for examples of massive goldstone fermions
and bosons, respectively, in the sense of  \cite{Watanabe:2013uya} in the context of the large charge expansion.). For purposes of counting light goldstone modes we can ignore these and count only symmetry generators that are spontaneously broken by the
solution at fixed time, and commute with the helical frequency.\footnote{This formulation of the goldstone-counting rule, while not manifestly equivalent to the way
of counting in \cite{Watanabe:2013uya}, is more convenient for our purposes and
does work out to the same answer, as illustrated
in \cite{Alvarez-Gaume:2016vff}.}

We will not refer further to the chemical potential in the present paper; we have included these comments only to allow the reader to translate
without difficulty between the two points of view.


\heading{The case of the $O(2)$ model}

The discussion so far has led us to a simple rule for the counting of light modes: The light modes that can be understood
as goldstone bosons, correspond one to one with symmetry generators that are spontaneously broken
by the configuration at fixed time $t=0$, and which commute with the generator $\hat{g}$ describing the helical
frequency.  As an example we will now apply this rule to the case of the large-charge ground state of the
$O(2)$ model.  We will see the rule gives a simple explanation for the spatial homogeneity of the ground state, that is independent
of the details of the Wilsonian action at the conformal fixed point.

In the case of the $O(2)$ model the nature of the large-charge ground state is by now well-understood:
In addition to the helical symmetry, the ground state is also spatially homogeneous, and the goldstone-counting 
rule shows this must be the case, because the $O(2)$ model does not have enough massless fields to
realize more than one goldstone boson.

The proof of the homogeneity goes as follows.
Assume otherwise, then the inhomogeneity is at some particular scale set by the charge density itself.
Then in the IR, the effective action should contain one or more translational Goldstone bosons and one axion from the spontaneous broken $O(2)$ (combined with broken time translation symmetry). 
Now, remember we started out from a theory of one complex scalar, with two real degrees of freedom.
The renomalization group flow takes this UV theory to an IR theory which inevitably includes fewer than two real light degrees of freedom, and hence contradicts with the above statement.
So by contradiction, we know that the ground state configuration must be homogeneous for the $O(2)$ WF fixed point at large charge.

One could have considered a logical possibility that the charged ground state may be inhomogeneous, and
indeed $O(2)$ or $U(1)$ symmetric CFT with a larger number of degrees of freedom, may spontaneously break
translational symmetries, because they have enough degrees of freedom from the start, to do so.
In the $O(2)$ case, the homogeneity of the ground state is related to the small number of light fields available,
and does not follow automatically from the symmetries of the conformal fixed point.

\heading{The case of the $O(4)$ model with only one nonzero charge}

Now we consider the next-simplest case, already analyzed in \cite{Alvarez-Gaume:2016vff}, and apply
the goldstone-counting argument to reproduce some results of that paper regarding the
spatial homogeneity of the ground state for various choices of ratios of charge density.  In \cite{Alvarez-Gaume:2016vff}
it was shown that a particular choice of (conjugacy classes of) large-charge limit, have a spatially homogeneous
ground state, namely those in which the charge density of the $SO(4) \simeq SU(2) \times SU(2)$ symmetry
of the CFT is chosen to lie entirely inside one of the two $SU(2)$ factors.  That is, letting 
$\rho\ll 1, \rho\ll 2$ be the Cartan eigenvalues of the total charge matrix (divided by a factor of the spatial volume), 
one can show that a classical ground state of the system with those charges, is homogeneous.

Here without loss of generality we set $\rho_2=0$, and let us use the notation $q$ as in \eqref{notation}.
In terms of $q$, this condition translates to, by looking at \eqref{charge} and \eqref{singlet}, $q_2\partial_t q_2=0$.
Just assume $q_2=0$ for the moment, and then the subgroup of $O(4)$ which preserves this condition is $SU(2)\times U(1)$ (overall phase rotation and $SU(2)$ rotation of $q_1$).

Because of how $q$ is defined, $q$ takes the following form,
\begin{equation}
q=
\begin{pmatrix}
q_1\\ q_2
\end{pmatrix}
=
\begin{pmatrix}
e^{i\omega_1 t}\sin(p(x))\\ e^{i\omega_2 t}\cos(p(x))
\end{pmatrix}.\llok
\end{equation}
So in order to have $q_2=0$, you inevitably have $p(x)=\pi/2$, so that the configuration is homogeneous.

Now consider the case where $\partial_t q_2=0$. The subgroup of $O(4)$ which preserves this condition is the same $SU(2)\times U(1)$ as before.
The solution to the equation of motion spontaneously breaks this $SU(2)\times U(1)$ into a smaller subgroup.
If we assume the ground state configuration is homogeneous, one of such solutions is
\begin{equation}
\partial_t q=
\begin{pmatrix}
1\\0
\end{pmatrix}
\end{equation}
but the subgroup of  $SU(2)\times U(1)$ which respects this particular ground state is only $U(1)$ phase rotation of $\partial_t q_2$.

So, schematically, we get the following breaking pattern;
\begin{equation}
O(4)\xrightarrow{\text{helical frequency}}	U(2)= {U}(1)\times {SU}(2)\xrightarrow{\text{spontaneous breaking}} U(1)
\end{equation}

The dimension of the coset is $\dim\left(U(2)/U(1)\right)=3$, so we have as many as 3 Goldstone bosons.
Note that this counting is precisely what is given in \cite{Alvarez-Gaume:2016vff}.
If you were to break the ground state homogeneity, you add one or more translational Goldstone bosons to these, making the total number of them 4 or more.
But you only have three light real degrees of freedom in the IR; thus by contradiction, the ground state configuration must always be homogeneous in the one-charge case.

\heading{The case of the $O(4)$ model with two nonzero charges}

In this case, we know from \cite{Alvarez-Gaume:2016vff} that the configuration is inhomogeneous, so let us first assume the configuration is inhomogeneous only in one direction of the torus.
As you still have the freedom to phase rotate $q_1$ and $q_2$, respectively
the pattern of breaking by the helical frequency becomes
 \begin{equation}
O(4)\xrightarrow{\text{helical frequency}}	{U}(1)\times {U}(1)
\end{equation}
 The inhomoegenity will not let us choose a ground state configuration which is special, like $\partial_t q=
\begin{pmatrix}
1\\0
\end{pmatrix}$.
Rather, the ground state configuration can only lie at a very generic point, hence preserves no subgroup of $U(1)\times U(1)$.
The spontaneous breaking pattern of it, along with translational symmetries, is therefore,
\begin{equation}
U(1)\times U(1)\times\{\text{translation}\}^2\xrightarrow{\text{spontaneous breaking}}	\{\text{translation}\}
\end{equation}
 so that you have three Goldstone boson in the system by looking at the coset dimension.
 
 Now, if you were to break one more translational symmetry, there would be four light Goldstone bosons in the system; however the
 theory has only had three light real fields with which to realize such excitations, which would be a
 contradiction, regardless of the form of the Wilsonian action for the light fields.
 Therefore the translational symmetry of the torus can be only broken in one direction for the case of the $O(4)$ model at generic set of large charges.

\section{The $O(4)$ model at arbitrary charge densities}
\label{review}

We move on to prove that the possibility (c-IR) is self-consistent
by extending the result in \cite{Hellerman:2017efx} to the case of any sets of charges, $J_{1,2}$.
In this paper, we denote by $J_{1,2}$ the two independent positive real eigenvalues of the matrix defined by the $SO(4)$ Noether charge.
Now, rather than requiring the limit where one of the global charge is much less than the other as was done in \cite{Hellerman:2017efx},
we study the ground state of the critical $O(4)$ model, fixing the
spatially averaged global
charge densities $\rho\ll{1,2} \equiv J\ll{1,2} / {\cal V}$ in an arbitrary
ratio $J\ll 1 / J\ll 2$.   
As in \cite{Hellerman:2017efx} we integrate
out the mode which becomes heavy at
large charge and work within a weakly-coupled conformal sigma model with target space $S\uu 3$.  
The Lagrangian density for this conformal effective theory is
singular in the vacuum but is not meant to be used there; it is only meant to be used to study states of energy $O(1)$ or less, above the ground state of large charge density. \ok

In \cite{Hellerman:2017efx}, we reproduced the no-go result of 
\cite{Alvarez-Gaume:2016vff} in the context of
the conformal sigma model: For generic charge densities or chemical potentials for the two independent $O(4)$ charges, there is 
no spatially homogeneous classical solution.  
The only
homogeneous ground states occur when 
the two chemical potentials are equal, or
equivalently when the real antisymmetric $4\times 4$ matrix defining the 
$O(4)$ charge of the state has a vanishing determinant, which in our conventions
and those of \cite{Alvarez-Gaume:2016vff},
means $\r\ll 1=0$ or $\r\ll 2= 0$. \ok

For any classical theory, the lowest energy
solution carrying given global charges
always leaves unbroken a helical 
symmetry, by which we mean a symmetry under combined time translation
and global symmetry rotation.
Regarding the spatial configurations are  
necessarily inhomogeneous in the $O(4)$ model with generic $O(4)$ charges, by virtue of \cite{Alvarez-Gaume:2016vff}, as we have stated earlier.
Hence in this note, we search for the lowest-energy inhomogeneous helical solution. \ok

\begin{sloppypar}
For inhomogeneous ground states, the most
important qualitative question is whether the
inhomogeneity is on the ultraviolet scale,
set by the charge density itself, or on the infrared
scale, set by the boundary conditions or
overall geometry and topology of the spatial slice.
It is only in the case where the inhomogeneity
is on the infrared scale, that the large-charge effective theory can be used in a straightforward manner.
\end{sloppypar} \ok

In the space of CFT with global symmetries, 
this question does not have a simplistic universal answer: The answer appears to depend on 
the theory and on the choice of global symmetry
quantum numbers.  For angular momentum
in a single plane, for example, the lowest
state is generally a small number of quanta
each carrying a large angular momentum
 \cite{Komargodski:2012ek,Fitzpatrick:2012yx} and thus
the inhomogeneity is on the UV rather than
IR scale.  \ok On the other hand, for
theories in $D\geq 4$ the ground state with
angular momentum in multiple planes, taken
large in fixed ratio, may have a smooth, semiclassical
ground state solution generically, which is
inhomogeneous on the infrared scale.  Both
these behaviours are visible even in free theories. \ok

For a given
set of global quantum numbers, then, the
behavior of the ground state is a question to
be settled by direct calculation.  For the case
of the O(4) model with generic $O(4)$ charges
$\r\ll{1,2}$ taken large in fixed ratio, it was shown in  \cite{Hellerman:2017efx} that the lowest state has inhomogeneity on the infrared scale, for $\r\ll 1 / \r\ll 2$ (or $\r\ll 2 / \r\ll 1$) sufficiently small.  In
the present paper, we show that this result holds
for any value of the ratio $\r\ll 1 / \r_2$, and
thus the ground state properties at large charge
can be analyzed consistently in the effective
conformal sigma model for any charge ratio. \ok

The simplest candidate helical solutions leave the translational symmetry unbroken in one spatial direction and break it in the other direction down to a discrete
subgroup whose period is $\ell$.
Two natural questions, which are closely related arise because of this fact.
One is which value of $\ell$ has the lowest energy.
The other one is the range of $\ell$ where the effective field theory is reliable. \ok\ok
We will now try to answer this question using effective field theory, and are going to show that
where $\ell\muchgreaterthan \sqrt{\rho^{-1}}$, the lowest energy is achieved where $\ell$ is largest, that is the size of the underlying geometry itself.

Note that it does {\it not} logically exclude the possibility of a striped phase, where the scale is set by $\sqrt{\rho^{-1}}$ itself so that the EFT breaks down.
Although this possibility could not be realized in the case of the $O(2)$ model due to Goldstone counting,
in the case of the $O(4)$ model we could not rule out this possibility on the basis of Goldstone-counting alone,
due to the fact that any value of $\ell$ would realize the same symmetry-breaking pattern, and so the Hamiltonian could a priori favor
either long or short distance scales for $\ell$.  The EFT analysis shows only that a solution with period $\ell$ that is long
on the infrared scale, energetically prefers as long a spatial period $\ell$ as possible, so that there is no internal instability
in the EFT towards inhomogeneity on short scales.

\subsection{Parametrizing the charge density}

A convenient parametrization for the $O(4)$ charge is as follows.
The charge densities take value in the adjoint of $O(4)$,
the group of antisymmetric $4\times 4$ complex matrices,
which has real eigenvalues that occur in pairs with equal magnitude
and opposite sign.
We denote two (out of four) positive eigenvalues of the charge density matrix to be
$J\ll 1$ and $J\ll 2$; in the infinite volume limit ${\cal V}\to
\infty$, we may take these to infinity, fixing the spatially averaged charge densities $\rho_{1,2} \equiv J\ll{1,2} / {\cal V}$. \ok

For helical solutions, this specific choice of parametrization is equivalent to
choosing a complex basis for the fundamental of $U(2)\subset SO(4)$
and parametrizing the charge generator by the two matrix elements on the diagonal.
Note that the off-diagonal matrix elements always vanish because the lowest-energy classical solution is always helical. 
This is the same convention used in the earlier work, \cite{Alvarez-Gaume:2016vff}. \ok

We are now going to put the system on $\mathbb{R}\times \mathbb{R}^2$
and seek for the ground state configuration.
This should be regarded as the infinite volume limit of the geometry $\mathbb{R}\times S^2$ or $\mathbb{R}\times T^2$.
We will comment on the ground state configurations in finite volume case later on.\ok

\subsection{Conformal sigma model from linear 
sigma model}

Now that the convenient parametrization of the charge densities has been made,
we describe the $O(4)$ model by four real scalars $X^{1,2,3,4}$,
which is then organised into $Q\equiv
 \begin{pmatrix}
X\ll 1 + i X\ll 2 \\ X\ll 3 + i X\ll 4
\end{pmatrix}$,
a complex $SU(2)$ doublet.
The interacting IR fixed point of the model is given by starting from the UV Lagrangian
with the kinetic term for $Q$ plus a quartic potential proportional to $|Q|^4$,
with a fine-tuned mass term so that it actually flows to a non-trivial fixed point. 
We parametrise $Q$ as amplitudes and angles, which is given by
\begin{equation}
Q=A\times q,\quad q=
\begin{pmatrix}
q_1\\ q_2 
\end{pmatrix},\llok
\label{notation}
\end{equation}
where $|q_1|^2+|q_2|^2=1$.
We give a large VEV to the $A$-field,
and the resulting leading action in the IR includes 
a term that is proportional to $A^6$, as explained in \cite{Hellerman:2015nra}.
The IR Lagrangian we get, henceforth, is
\begin{equation}
\mathcal{L}_{\mathrm{IR}}=\frac{1}{2}(\partial A)^2+\frac{\gamma}{2}A^2\partial q^\dagger \partial q-\frac{h^2}{6}A^6.
\label{IRLagrangian0}\llok
\end{equation}
This is under a RG normalization condition that the two-derivative kinetic term of $A$ is canonical. \ok

Note that we have omitted other terms like Ricci coupling and higher derivative terms, because we are only using the leading large-charge-density term.
We already know that there are no homogeneous ground state configurations, so
the suppression of these terms, however, should be proven {\it a posteriori}. 
Those terms are only suppressed when the scale of the inhomogeneity, $L$, is much larger than the UV scale, $(\r\ll 1 + \r\ll 2)\uu{-\hh}$ --
otherwise the large-charge EFT is not within its range of validity, and the higher-derivative operators and quantum corrections are out of control, and
there is no simplification of the dynamics at large charge densities.
 In the following we first assume $L$ to be much smaller than the UV scale and derive the ground state configuration, and then justify this assumption later,
 {\it a posteriori}.\llok

Under this assumption, we first integrate the $A$ field out, which has a mass scale
defined by the charge density itself, which is above the Wilsonian scale we are talking about.
By virtue of the EOM for $A$, we have, as an equilibrium value of $A$, 
\begin{equation}
\frac{\delta \mathcal{L}_{\rm IR}}{\delta A}=0
\iff A^2=\sqrt{\frac{\partial q^\dagger \partial q}{\gamma^{-1} h^2}}\llok
\end{equation} 

Now by using this and the original IR Lagrangian, \eqref{IRLagrangian0},
we get
\begin{equation}
\mathcal{L}=b_q\mathcal{L}_0^{3/2}=b_q(\partial q^\dagger \partial q)^{3/2},\llok
\label{conformalsigma}
\end{equation}
which is the conformal sigma model whose target space is $S^3$ and
where $|q| = 1$ and $b_q=\sqrt{\gamma^3h^{-2}}/3$ is an undetermined coefficient from the original large-charge effective action \cite{Hellerman:2015nra}.\ok

\subsection{Restriction to fixed average charge densities $\r\ll{1,2}$  }

We now put the theory on $\mathbb{R}^2$, so we inevitably have to
use the concept of the ``fixed average charge density'' instead of that of total charge, which is ill-defined.
We therefore impose the following conditions onto Noether currents,
\begin{eqnarray}
-\left.\frac{2ib_q}{3}\int dx^{i} \sqrt{\mathcal{L}_0}\left[q^{\dagger}\partial_t q {{{}-\mathrm{{}c.c.}}}\right]\right/\mathcal{V}&=&\rho_1+\rho_2\label{charge}
\\
-\left.\frac{2ib_q}{3} \int dx^{i} \sqrt{\mathcal{L}_0}\left[q^{\dagger}\sigma^{3}\partial_t q{{{}-\mathrm{{}c.c.}}}\right]\right/\mathcal{V}&=&\rho_1-\rho_2, 
\label{singlet}
\end{eqnarray}
where $\mathcal{V}$ indicates the total volume of the space. Also, let us set $\rho_1<\rho_2$ for simplicity, but we will comment on the $\rho_2<\rho_1$ case later on.
\llok

The energy density can also be derived from the Lagrangian, which is
\begin{equation}
\mathcal{H}=
 b_q\sqrt{\dot{q}^\dagger\dot{q}-\partial_i q^\dagger \partial^i q}\times \left(2\dot{q}^\dagger\dot{q}+\partial_i q^\dagger \partial^i q\right).
 \label{energydensity}
\end{equation}
We will therefore look for the minimizer of above under constraints
\eqref{charge} and \eqref{singlet}. \llok

\subsection{Equation of motion for the conformal sigma model}\label{MWSigmaModelEOM}

To achieve the ground state solution of \eqref{conformalsigma},
we set an ansatz that the solution is at least homogeneous in one of the spatial directions, $y$, and only varies spatially in the $x$ direction.

We also use the helical nature of the ground state solution
and the invariance under the combination of $t\to -t$ and the complex conjugation.
Basically these ansatz sets the solution of the form
\begin{equation}
q=
\begin{pmatrix}
q_1\\ q_2
\end{pmatrix}
=
\begin{pmatrix}
e^{i\omega_1 t}\sin(p(x))\\ e^{i\omega_2 t}\cos(p(x))
\end{pmatrix},\llok
\end{equation}
where we are free to set $\omega_1>\omega_2$
and $p(x)$ takes value in $\mathbb{R}$.
Under this parametrization, \eqref{conformalsigma} becomes
\begin{equation}
\mathcal{L}=b_q\left[-p^\prime(x)^2+V(p)\right]^{3/2},
\llok
\end{equation}
where 
\begin{equation}
V(p)=\omega_2^2+\left(\omega_1^2-\omega_2^2\right)\sin^2(p).\llok
\label{VFormula}
\end{equation}

For general helical solutions depending on no more than one
spatial direction, we can simplify the equations by
reducing them to first order.  Local conservation of momentum
in the $x$-direction implies $0 = \pp\ll\m T\ll x {}\uu\m$.  For a helical
solution independent of the $y$-direction, the stress tensor is
independent altogether of $y$ and $t$, so the pressure $T\ll{xx}$ in
the $x$-direction is simply a constant:
\bbb
\pp\ll \m T\ll{xx} = 0\ .\llok
\eee
Define a constant $\kappa$, of mass dimension $+1$ as
the cube root of the pressure, with the coefficient of
proportionality 
$b\ll q\uu{-{1\over 3}} $ to simplify the formulae:
\bbb
\k \equiv  
b\ll q\uu{-{1\over 3}} \cc T\ll{xx} \uu{{1\over 3}}\ .\llok
\eee

Now we will use the general formula for the stress tensor in a theory with a
Lagrangian that is first-order in derivatives acting on fields of vanishing
conformal weight:
\bbb
T\ll\m{}\uu\n \equiv \d\ll\m{}\uu\n \cc {\cal L} - \sum\ll A
\Theta\uu A\ll{,\m}{{\d {\cal L}}\over
{\d \Theta\uu A\ll{,\n}}}\llok
\een{StressTensorFormula}
where $\Theta\uu A$ runs over all the fields in the system, in this case
the three Goldstones parametrizing the $S\uu 3$ target space.
For a helical solution, with $\dot{\chi}\ll{1,2}
= \o\ll{1,2}$, the Lagrangian density is
\bbb
{\cal L} = b\ll q \cc {\cal L}\ll 0\uu{3\over 2}\ ,\llok
\xxx
{\cal L}\ll 0 \equiv V - p\prsqd\ ,\llok
\xxx
V\equiv \o\ll 2\sqd +\left (\o\ll 1 \sqd - \o\ll 2 \sqd\right)\cc  \sin \sqd(p)\ .\llok
\eee
modulo terms of second order in $\chi\pr\ll{1,2}$, which do not
contribute to the stress tensor in a helical solution because
$\chi\pr\ll{1,2}$ vanishes in the helical solution itself.

So the stress tensor is
\begin{equation}
T\ll{xx} = \mathcal{L}-p^{\prime}(x)\frac{\delta \mathcal{L}}{\delta p^{\prime}}=b_q
\sqrt{-p^{\prime}(x)^2+V(p(x))}
\left(2p^{\prime}(x)^2+{V(p(x))}\right),\llok
\end{equation}
which we know is a constant and have already set to be equal to $b_q\kappa^3$.
We now have the EOM, which is
\begin{equation}
-\frac{\kappa^6}{4}=-\frac{b_q^{-2}T_{xx}^2}{4}=\left(p^{\prime}(x)^2-V(p(x))\right)
\left(p^{\prime}(x)^2+\frac{V(p(x))}{2}\right)^2,\llok
\label{FirstOrderEOM}
\end{equation}
where $\kappa>0$ because of the positivity of $T_{xx}$.
The meat of this is that the equation of motion has now been reduced to a first-order
equation \rr{FirstOrderEOM} with one undetermined constant of
motion, $\k$.  In principle we could invert \rr{FirstOrderEOM} 
algebraically to solve for $(p^\prime)^2$ in terms of $V(p)$ for a given
$\k$, using Cardano's formula for the general solution to a cubic
equation, and then solve the first-order autonomous ODE for
$p$ as a function of $x$.  However most of the complication involved
in such a solution is unnecessary, because we are working only
within the regime where the fields are varying on scales
$L$ long compared to the ultraviolet scale set by $\o\ll{1,2}$,
so we need only solve the EOM under the condition
\bbb
p\pr \muchlessthan \o\ll{1,2}\ .\llok
\een{LowEnRegime}
Indeed, there are other terms in the effective Lagrangian that we have
omitted, which would become important if we were to work outside this
regime.  But now we will now organize the first-order EOM in such a way as 
to exploit the condition \rr{LowEnRegime} in order to solve it.
\llok

\heading{Remark}

Before proceeding, let us make a few comments about how to choose the right solution when solving \eqref{FirstOrderEOM} for $(p^\prime)^2$.
By imposing $(p^\prime)^2>0$ and $\kappa>0$,
we can have multiple solutions depending on the value of $\kappa$.
Here, however, we only restrict attention to the case where $0\leqslant (p^\prime)^2\leqslant V(p)/2$. 
This is equivalent to imposing a condition that $p(x)$ must have a point where its derivative is vanishing.
This is natural when we eventually want to put the system on $S^2$ and compute the dimensions of operators using the Neumann boundary condition on the poles.
When we put the system on a torus, $T^2$, there can possibly be solutions
on different ``winding numbers'', that are characterised by $p(x/\ell)=p(0)+n\pi$.
The winding number here is not a topological charge, because the map from
$T^2$ to $S^3$ can only be trivial homotopically.
This means that on the torus we will also have to compute their energies
separately, to know the true ground state configuration.
We will, however, just assume
the lowest solution is achieved when $n=0$ even on the torus --
at least we know that this is the case in the homogeneous case, and the continuity requires the statement is also true in a certain subset of $J_1/J_2$ near zero.
Also, this is physically related to the existence of soft modes discussed in \cite{Alvarez-Gaume:2016vff}, and adding soft modes intuitively should increase the energy of the system, not otherwise.
Again for these reasons, we will hereafter only consider solutions which has a point at which its derivative vanishes.
\ok

\heading{Scales in the equation of motion}

The dimensionful quantities in the EOM are $p\pr, \o\ll 1$ and $\o\ll 2$,
and we are going to consider the regime \rr{LowEnRegime}.  The two
frequencies $\o\ll{1,2}$ are independent a priori, but their relationship
will be fixed in terms of the spatial period of the solution.\ok

First thing to notice is that the spatial period of the solution goes to infinity at $\o\ll 1 = \o\ll 2$, becuase the EOM just gives $p'(x)=0$.
 This tells us straight away that the
difference in frequencies, $\o\ll 1 - \o\ll 2$ must scale differently 
than either individual frequency: while we can hold
$\o\ll 1$ or $\o\ll 2$ fixed while taking $p\pr \to 0$, 
we see that $\o\ll 1 - \o\ll 2$ must vanish in the limit $p\pr \to 0$.
Let us now rewrite
the first order EOM \rr{FirstOrderEOM} to emphasize 
the distinction in scalings.  We define $\o\ll - \equiv \o\ll 1 - \o\ll 2$ and
expand $\k$ as $\k\ls 0 + \D\kappa$, where $\k\ls 0$ is
the value of $\k$ for a homogeneous solution with a given $\o\ll 2$,
and $\D \kappa \equiv \k - \k\ls 0$ is the difference, which 
must scale as a positive power of $\o\ll -$.  We could further expand
$\D\k$ as a series $\k\ls 1 + \k\ls 2 + \cdots$ where $\k\ls p$ is
the term of order $\o\ll -\uu p$; this will not be necessary, however, as
we will only be interested in first-order quantities. 
So using the formula \rr{VFormula}, which we recap here,
\begin{equation}
V(p)=\omega_2^2+\left(\omega_1^2-\omega_2^2\right)\sin^2(p)\ ,\llok
\label{VFormulaRECAP}
\end{equation}
we see that $V = O(\o\ll 2\sqd) + O(\o\ll 2\o\ll -)$.  So $V\uu 3$ is
of order $O(\o\ll 2\uu 6) + O(\o\ll 2\uu 5 \o\ll -)$.  So now, the LHS
of \rr{FirstOrderEOM} is identically independent of spacetime;
therefore the $x$-dependent parts of the RHS will have
to cancel order by order in $\o\ll - / \o\ll 2$.  \ok

The two types of $x$-dependent terms on the RHS 
are $p\prsqd \o\ll 2\uu 4$ and $\o\ll - \o\ll 2\uu 5 \cc  \sin \sqd(p\ll 0)$.  In order for them to cancel, if one is treating $p\ll 0$ as $O(1)$, one needs to scale 
$\o\ll -$ as $p\prsqd / \o\ll 2$.  \ok

To make this more concrete, we want to define the length scale $\ell$ as the inverse
of the maximum $x$-derivative of the $p$-field.  This isn't quite the
right thing, because we want to get the right order of magnitude
for the length scale not only when the amplitude of the oscillations is of
$O(1)$ but also when it is small, whereas assigning $\ell$ to be ${1\over{p\pr\lrm{max}}}$ would go to infinity in the limit when the amplitude of the oscillations
is small but the period is fixed.  To repair this deficiency, we multiply by
$\sin p\ll 0$, and define:
\bbb
\ell \equiv {{ \sin (p\ll 0)}\over{p\pr\lrm{max}}}\ .\llok
\een{GeneralScaleEllDef}
We will express $\ell$ in terms of the actual period of the solution later.

This defines the general length scale characterizing the solution, and
we expect the actual spatial period of the solution to be of order $\ell$;
we will confirm this expectation when we find the ground-state classical solution exactly.\ok

In terms of $\ell$, then, we see $\o\ll -$ must scale as
\bbb
\o\ll - = O \cc \big ({1\over{\ell\sqd \o\ll 2}} \big )\ .\llok
\eee
We require that there be a point at which $p\pr$ vanishes, and at that point
we have to be able to satisfy the EOM anyway.  So this fixes
$\Delta \k$ at first order completely in terms of $\o\ll 2, \o\ll -$ and $p\ll 0$:
\bbb
\Delta \k = \k\ls 1 + O\big ( \cc \ell\uu{-4} \o\ll 2\uu{-3} \cc \big )\ ,\llok
\xxx
\k\ls 1 = \o\ll -\cc  \sin \sqd (p\ll 0) \ , \llok
\eee
which we can also write as
\bbb
\k\ls 1 = \o\ll 2 \cc \eta\ ,\llsk\llsk 
\eta\equiv {{\o\ll -}\over{\o\ll 2}}\cc  \sin \sqd(p\ll 0)\ .
\llok
\eee
Now, at first order in $\o\ll -$, the EOM reads:
\bbb
p\prsqd  = 2\o\ll 2\o\ll - \cc \big [ \cc  \sin \sqd (p\ll 0) -  \sin \sqd(p) \cc \big ] \ ,\llok
\een{FirstOrderEOMAtFirstOrder}
which we can also write as
\bbb
p\prsqd = 2\o\ll 2\sqd \big [ \cc \eta - \e\cc  \sin \sqd(p) \cc \big ]\ ,
\llsk\llsk \e \equiv {{\o\ll -}\over{\o\ll 2}}\ .\llok
\een{MasatakaStyleEOM}
The quantities $\e$ and $\eta$ are related simply by 
\bbb
\eta =  \sin \sqd(p\ll 0) \e\ ,\llok
\eee
so the ratio $\eta / \e$ is always less than $1$, and goes to zero
in the linearized limit, where the amplitude of oscillation is small.

The maximum value of $p\pr$ occurs when $p=0$, where it takes the value
\bbb
(p^{\prime})\lrm{max} = \sqrt{2\o\ll 2 \o\ll -} \cc | \sin (p\ll 0)| = \sqrt{2\e} \cc \o\ll 2\cc
 | \sin (p\ll 0)| = \sqrt{2\eta}\cc \o\ll 2\ , \llok
\een{PPrimeMaxFormula}
and so the general scale $\ell$ of the solution, which we defined in
\rr{GeneralScaleEllDef}, is
\bbb
\ell = {1\over{\sqrt{2\o\ll 2\o\ll -}}} = {1\over{\sqrt{2\e}\cc \o\ll 2}}\ .\llok
\een{GeneralScaleEllFormula}

This low-energy EOM \rr{FirstOrderEOMAtFirstOrder} for the helical solution
is exactly the equation of motion for the angle ${\boldsymbol\theta}$ of the pendulum
with length ${\bf L}$ in a gravitational field ${\bf g}$ under the identification of $p={\boldsymbol\theta}/2$ and $2\omega_2\omega_{-}={\bf g}/{\bf L}$
We know the exact solution to this type of differential equations so
now we can simply solve the equation \rr{FirstOrderEOMAtFirstOrder}
and using it, we can compute the energy of the large charge ground state.\ok

Now we have the analytic solution to \rr{FirstOrderEOMAtFirstOrder}, which is
\def\sn{\mathop{\rm sn}}
\begin{equation}
{{ \sin (p)}\over{ \sin (p\ll 0)}} = 
 \sn \left (\cc {x\over\ell}\cc \cc ; \cc \cc   \sin (p\ll 0)\cc 
\right )\llok
\label{ExactSolution}
\end{equation}
where $\ell$ was already given in \rr{GeneralScaleEllFormula}, and $\sn(x;k)$ is the Jacobian elliptic function with modulus $k$.
The quarter period of the solution, $L$, is given by
\begin{equation}
L =
\ell\cc F\left ( \cc \frac{\pi}{2} \cc ; \cc  \sin (p\ll 0)\cc \right )
\llok
\label{GeneralScaleToQuarterPeriodTranslation}
\end{equation}
where 
\bbb
F \left ( \cc p\cc ; \cc k \cc \right ) \equiv \int_0^p {{ {d\hat{p}}}\over{{\sqrt{1-k^2\sin^2(\hat{p})}}}}={\sn}^{-1}\left ( \cc \sin(p);k \cc \right )\cc\cc\cc \llok
\een{CapitalFDef}

\heading{Fixing the charge densities}

We have now solved for the spatial dependence $p(x)$ in terms of the amplitude $p\ll 0$ and 
the frequencies $\o\ll{1,2}$.  However we do not really want $\ell$ to be an
output and we do not really want the spatial frequencies to be inputs.
We would like to invert the relationship between the frequencies $\o\ll{1,2}$
and the densities $\r\ll{1,2}$ so that the latter are the independent
variable.

The expressions for the average
charge densities given by \eqref{charge} and \eqref{singlet}, are
\begin{eqnarray}
\rho_1&=&\frac{8b_q}{3\mathcal{V}}\int dx^i\, \omega_1\sqrt{-p^{\prime}(x)^2+V(p(x))}\sin^2(p(x))
\llok
\label{baryonsin}\label{Rho1Formula}\\
\rho_2&=&\frac{8b_q}{3\mathcal{V}}\int dx^i\, \omega_2\sqrt{-p^{\prime}(x)^2+V(p(x))}\cos^2(p(x)),
\llok
\label{baryoncos}\label{Rho2Formula}
\end{eqnarray}

We can compute these integrals analytically at leading-order in the low-energy
expansion $\ell\muchgreaterthan 1/\omega_2$.
Using the equation of motion \rr{GeneralScaleEllFormula}, we can substitute
\begin{equation}
\sqrt{-p^{\prime}(x)^2+V(p(x)))}= \frac{1}{\sqrt{2\eta}}\frac{p^{\prime}(x)(1+\eta)}{\sqrt{
{1-\frac{\epsilon}{\eta}\sin^2(p(x))}}}+O(\epsilon^2).
\llok
\label{fo}
\end{equation}

Now as the previous subsection essentially states that the low-energy expansion we are interested in is in terms of $\epsilon,\, \eta\muchlessthan 1$ where $\eta/\epsilon=\sin^2(p_0)=O(1)$, we only have to keep track of the leading order contribution in terms of this expansion rule.
\eqref{fo}, therefore means that at first order in $\epsilon$ and $\eta$,
we can now change the variable in the integrand from $x$ to $p$ itself.
Doing this, and using $\eta / \e =  \sin \sqd(p\ll 0)$, we have
\begin{eqnarray}
\rho_1&=&
\frac{8b_q\omega_2(1{+\epsilon})(1+\eta)}{3L\sqrt{2\eta}}\int\limits_0^{p_0}dp\,
\frac{\sin^2(p)}{\sqrt{
{1-{{\sin^2(p) }\over{\sin\sqd(p\ll 0)}}}}}
\llok
\\
&=&
 \frac{8b_q\omega_2^2(1{+\epsilon}+\eta)}{3}\times\Delta\left(\frac{\pi}{2};\cc
 \sin (p\ll 0)\right )
 \llok
\label{ExpressionThatEqualsTheOneBelow}\\
\rho_2&=& 
\frac{8b_q\omega_2^2(1+\eta)}{3} \cc\left [ \cc 1 - \Delta\left(\frac{\pi}{2};\cc
 \sin (p\ll 0)\right ) \cc\right ]
 \llok
 \label{rho2}
\end{eqnarray}
where 
\bbb
\Delta\left(p \cc ; \cc k\right )
\equiv
\frac{F\left(p \cc ; \cc k\right )-E\left(p \cc ; \cc k\right )}{F\left(p \cc ; \cc k\right )}\llok
\xxx
F(x;k)\equiv \int_0^p 
{{
{d\hat{p}}
}
\over
{\sqrt{1-k^2\sin^2(\hat{p})}}
}\llok
\xxx
E(p;k) \equiv \int_0^p {d\hat{p}}{\sqrt{1-k^2\sin^2(\hat{p})}}
\llok
\een{DeltaAndCapitalEDefPRECAP}

These expressions are a bit complicated, and it is instructive
to see how they behave in the linearized limit $p\ll 0 \muchlessthan 1$,
using the fact $\Delta\left(\frac{\pi}{2} \cc ; \cc \sin (p\ll 0) \right)\sim {{ \sin \sqd(p\ll 0)}\over 2} \sim
{{p\ll 0\sqd}\over 2}$ at leading order in $p\ll 0$.
We can also immediately reproduce the statement that limit $p_0 \muchlessthan 1$
and the limit $J_1\muchlessthan J_2$ is the same thing.

\heading{Average energy density}

The average energy density for the ground state configuration can be
calculated from \eqref{energydensity} as
\begin{eqnarray}
\mathcal{E}&=&\frac{b_q}{\mathcal{V}}
\int d^2x\, \sqrt{-p^{\prime}(x)^2+V(p(x))}(p^{\prime}(x)^2+2V(p(x))) \llok
\\
&=&\frac{b_q\kappa^3}{2}+\frac{9}{16}(\rho_1\omega_1+\rho_2\omega_2).\llok
\end{eqnarray}
Now by using
\begin{equation}
(\omega_2)^2=\frac{3(\rho_1+\rho_2)}{8b_q}\left(1-\eta-\frac{\epsilon\rho_1}{\rho_1+\rho_2}\right),\llok
\end{equation}
we get
\begin{equation}
\mathcal{E}
=\frac{3\sqrt{3}}{8\sqrt{2b_q}}(\rho_1+\rho_2)^{3/2}\times\left(
1+\frac{\eta}{2}+\frac{\epsilon\rho_1}{2(\rho_1+\rho_2)}
\right) \llok
\end{equation}
or you could also write this way using $\eta=\epsilon\sin^2(p_0)$,
\begin{equation}
\mathcal{E}
=\frac{3\sqrt{3}}{8\sqrt{2b_q}}(\rho_1+\rho_2)^{3/2}\times\left[
1+\frac{\eta}{2}\left(1+\frac{\rho_1}{\sin^2(p_0)(\rho_1+\rho_2)}\right)
\right] \llok
\label{ennnn}
\end{equation}
Now we also have the constraint due to fixed charges, which is
\begin{equation}
\frac{\rho_1}{\rho_1+\rho_2}=\Delta\left(\frac{\pi}{2};\cc
 \sin (p\ll 0)\right )\left(1+\epsilon\times\Delta\left(\frac{\pi}{2};\cc
 \sin (p\ll 0)\right )\right)\llok
\end{equation}
or equivalently,
\begin{equation}
\Delta\left(\frac{\pi}{2};\cc
 \sin (p\ll 0)\right )
 =\frac{\rho_1}{\rho_1+\rho_2}\left(1-\frac{\epsilon\rho_2}{\rho_1+\rho_2}\right), \llok
 \label{sin}
\end{equation}
which determines $\sin(p_0)=\sqrt{\eta/\epsilon}$.
We also have, from the constraint on the charges,
\begin{equation}
\rho_2=
\frac{4b_q\sin^2(p_0)}{3\ell^2} \cc\left [ \cc 1 - \Delta\left(\frac{\pi}{2};\cc
 \sin (p\ll 0)\right ) \cc\right ]\times \left(1+\frac{1}{\eta}\right)
 \llok
\end{equation}
hence
\begin{equation}
\eta=\left[\frac{3(\rho_1+\rho_2)\ell^2}{4b_q\sin^2(p_0)}-1\right]^{-1}
=\frac{4b_q\sin^2(p_0)}{3(\rho_1+\rho_2)\ell^2}\left(1+O(1/\ell^2)\right)
\llok
\end{equation}

The energy density, in terms of the length scale $\ell$ is, therefore,
\begin{equation}
\mathcal{E}
=\frac{3\sqrt{3}}{8\sqrt{2b_q}}(\rho_1+\rho_2)^{3/2}\times\left(
1+\frac{A}{\ell^2}
\right),\llok
\label{enfin}
\end{equation}
where 
\begin{equation}
A\equiv \frac{2b_q}{3(\rho_1+\rho_2)}\left(\sin^2(p_0)+\frac{\rho_1}{\rho_1+\rho_2}\right)>0. \llok
\label{A}
\end{equation}
(This reproduces the result in \cite{Hellerman:2017efx} for $\rho_1/(\rho_1+\rho_2)\to 0$.)
Therefore,
the ground state configuration is achieved when $L=\ell F\left(\frac{\pi}{2};\cc
 \sin (p\ll 0)\right )\to\infty$, which is the same result we have already got for $p\ll 0\muchlessthan 1$ in \cite{Hellerman:2017efx}.

\heading{Two branches of the solution}
We have worked
in the regime where $\rho_1<\rho_2$, but
what happens if we make $\rho_1>\rho_2$?
Does $\sin(p_0)$ go very close to $1$ and as a consequence?
The answer is no: one immediate reason is that the expression for the energy must be symmetric under the exchange of $\rho_1$ and $\rho_2$.
We could also say that the exchanging the role of them when $\rho_1>\rho_2$ is energetically favourble. 
Hence, when $\rho_1$ exceeds the value of $\rho_2$, there is a first order phase transition (strictly only a crossover but gets infinitely sharp at large-$J$ limit) and $\sin(p_0)$ decrease as $\rho_1/\rho_2$ gets bigger, eventually reaching the same homogeneous solution, $p_0=0$.

\section{Observables}
\label{main}

We compute various quantities for the $D=3$
Wilson-Fisher conformal $O(4)$ model on $T^2\times \mathbb{R}$ in this section.
Here we set $\ell_2<\ell_1$ to be the spatial period of the torus, the spatial slice.
Imposing the boundary condition on the toric geometry, then, we have $4\ell F\left(\frac{\pi}{2};\cc
 \sin (p\ll 0)\right )=4L=\ell_1$
or equivalently, $\ell=\frac{\ell_1}{4F\left(\frac{\pi}{2};\cc
 \sin (p\ll 0)\right )}$
\subsection{Energy}

Let us compute the total energy of the system when $\rho_1=\rho_2=\rho/2$ for definiteness.
According to \eqref{enfin}, the energy density is given by
\begin{equation}
\mathcal{E}
=\frac{3\sqrt{3}}{8\sqrt{2b_q}}\rho^{3/2}\times\left(
1+\frac{A}{\ell^2}
\right),\llok
\end{equation}
where 
\begin{equation}
A=\frac{b_q}{3\rho}\left(\sin^2(p_0)+\frac{1}{2}\right)
\llok
\end{equation}
and
\begin{equation}
\Delta\left(\frac{\pi}{2};\cc
 \sin (p\ll 0)\right )=\frac{1}{2}\iff \sin^2 (p\ll 0)=0.826\dots.
 \llok
\end{equation}
Using the value for $\sin^2(p_0)$, we also have
\begin{equation}
F\left(\frac{\pi}{2};\cc
 \sin (p\ll 0)\right )=2.32\dots.
 \llok
\end{equation}
To sum up, the total energy of the system becomes
\begin{equation}
E=\ell_1\ell_2\mathcal{E}
=\frac{3\sqrt{3}\ell_1\ell_2\rho^{3/2}}{8\sqrt{2b_q}}\times\left(
1+\frac{114.2\times b_q}{3\rho(\ell_1)^2}\right),
\llok
\end{equation}
or using $J\equiv\rho\ell_1\ell_2$,
\begin{equation}
E=\frac{3\sqrt{3}J^{3/2}}{8\sqrt{2b_q\ell_1\ell_2}}\times\left(
1+\frac{114.2\times b_q\ell_2}{3J\ell_1}\right).
\llok
\end{equation}
This shows that the classical ground state energy of the $O(4)$ model at large charge $J_1=J_2$ scales as 
$(J_1+J_2)^{3/2}$.
There is also the subleading term scaling $O(J^{1/2})$ here and this is also uncorrected by the quantum corrections on the torus.
This is because the only effective operator scaling as $O(J^{1/2})$ is $\mathtt{Ric}_3|q|^2$, which vanishes on the torus.\ok

\subsection{Two-point functions}

Let us calculate the two point function $\braket{q_1^{*}(0,0)q_1(x_1,x_2)}$.
At leading order there are just classical contributions, which amount to
\begin{equation}
\braket{q_1^{*}(0)q_1(x)}
\propto\int dy_1 dy_2 \, q_1^{*}(y_1,y_2)q_1(x_1+y_1,x_2+y_2) 
\end{equation}
because of the translational symmetry.
Then we have
\def\sn{\mathop{\rm sn}}
\def\cn{\mathop{\rm cn}}
\def\dn{\mathop{\rm dn}}
\begin{eqnarray}
\braket{q_1^{*}(0)q_1(x)}
&=&\int dy_1 dy_2 \, q_1^{*}(y_1,y_2)q_1(x_1+y_1,x_2+y_2)
\llok
 \\
&\propto&\int dy_1 \, \sin(p(y_1))\sin(p(x_1+y_1))
\llok
\\
&=&\sigma^2\int\limits_0^{4L} dy_1 \, 
\sn \left(\frac{y_1}{\ell};\sigma\right)
\sn \left(\frac{x_1+y_1}{\ell};\sigma\right)\llok
\end{eqnarray}
where $\Delta(\pi/2;\sigma)=\rho_1/(\rho_1+\rho_2)$.
Likewise, we have
\begin{eqnarray}
\braket{q_2^{*}(0)q_2(x)}
&\propto&\int\limits_0^{4L} dy_1 \, 
\dn \left(\frac{y_1}{\ell};\sigma\right)
\dn \left(\frac{x_1+y_1}{\ell};\sigma\right)\llok
\end{eqnarray}
Note that we have kept all the $\sigma$ dependence to check the result in the homogeneous limit, $\sigma\to 0$.
Also incidentally, because of the charge conservation, $\braket{q_1^{*}(0)q_2(x)}$
must vanish.

\begin{figure}[htbp]
  \begin{center}
    \includegraphics[width=0.7\columnwidth]{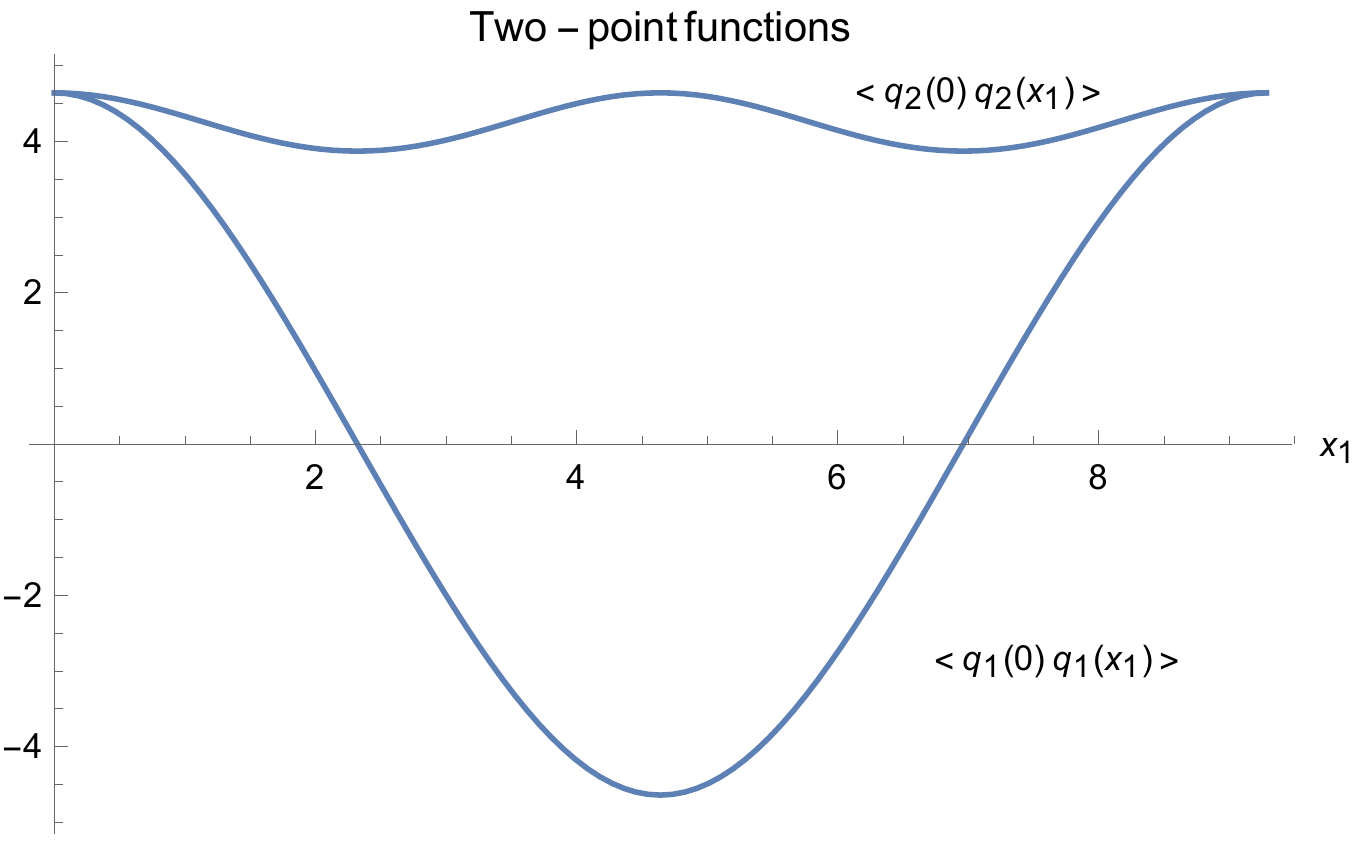}
    \caption{$2\pi\braket{q_1^{*}(0)q_1(x_1)}$ and $2\pi\braket{q_2^{*}(0)q_2(x_1)}$ in terms of $x_1$
when $J_1=J_2$. 
Note that they coincide with each other at $x_1=0$.}
    \label{2point}
  \end{center}
\end{figure}

These integrals can be performed analytically too, but this gives rather involved expressions involving compositions of
elliptic integrals and their inverses, and derivatives
of those.  It is simpler for practical purposes to express
the observables in terms of numerically evaluated
integrals; as an illustration, we compute these integrals numerically when $\rho_1=\rho_2$.
We also set $\ell=1$ with no loss of generality, since
the theory is conformal.
Note that the constraint $\Delta(\pi/2;\sigma)=\rho_1/(\rho_1+\rho_2)=1/2$ is equivalent to
$\sigma=0.91\dots$. 
Now by using this, we have the graph of the two-point function below (Fig.~\ref{2point}).
Meanwhile in the limit where $\rho_1\to 0$ the solution is just $p(x)=0$ and
we just get $\braket{q_1^{*}(0)q_1(x_1)}=0$ and $\braket{q_2^{*}(0)q_2(x_1)}=1$.\ok\ok

\subsection{Cluster decomposition and the infinite volume limit}

We have seen that the large-charge ground state of the system with a generic charge ratio $J\ll 2 / J\ll 1$,
is inhomogeneous, and that the inhomogeneity is energetically favored at the longest possible distance scales.
In finite volume, with a toroidal spatial slice, this pattern of global and translational symmetry breaking, produces
a distinctive signature in the correlation functions in the large-charge ground state.  Although our focus is
on finite-volume observables,  it is worth trying to understand the meaning of the infrared-enhanced inhomogeneity
in the infinite-volume limit.

Since the inhomogeneity is
clearly relevant at the longest distance scales,
these IR-inhomogeneous
solutions, may be understood as realizing a disordered phase of some kind, in which certain local operators
fail to cluster.  This is distinct from a striped phase,
in which the inhomogeneity would have a characteristic scale , which is inevitably fixed by the average charge density itself, in the case 
of a conformal field theory such as the critical $O(4)$ model. 

The phenomenon of violation of cluster decomposition is quite common under renormalization group flow.
What is unfamiliar here is the direct visibility of cluster-nondecomposition in 
large-charge perturbation theory, despite the strong coupling of the underlying model, where the perturbation parameter is $1 / (\mu |x|)$ , with $|x|$
being the distance between operators and $\mu$ being the chemical
potential.  The
non-clustering of the two-point function comes from
averaging over classical solutions that break the translational
invariance spontaneously at the IR scale.  Other examples of perturbatively calculable breakdown of cluster decomposition
under renormalization group flow are known, 
and are related by dualities to cases where defect operators fail to cluster due to strong coupling effects: For instance, see \cite{Hellerman:2006zs}, \cite{Caldararu:2007tc}, \cite{Hellerman:2010fv}.

\section{Results and discussions}

In this paper we began with a general argument, by counting the number of Goldstones and matching with the number of avaliable light modes, that for the $O(2)$ model at large global charge, the ground state configuration can only be homogeneous regardless
of the detailed form of the Wilsonian action at the fixed point.
We went on to extend the argument to give similar constraints for the ground state of the $O(4)$ model, especially that there can be inhomogeneity in only one direction in the case of generic charge densities.

We have also followed up on the argument due to \cite{Alvarez-Gaume:2016vff} that for the $O(4)$ model with a generic set of fixed large charges, the lowest energy classical solution is inhomogeneous: Making the argument more concrete, we have constructed the inhomogeneous ground state solution explicitly.
We also see that there are two branches of the configuration, namely when $J_1< J_2$ and when vice versa;
there is a continuous but nondifferentiable dependence
of observables on the charge ratio, at the point $J\ll 2 / J\ll 1 = 1$.  This solution confirms the analysis of \cite{Hellerman:2017efx}
at charge densities close to the homogeneous case, $J\ll 2 / J\ll 1 \muchlessthan 1$.

We have computed the leading order energy on he torus and saw that it scales as $(J_1+J_2)^{3/2}$ at leading order
just like the $O(2)$ model.
We computed the two-point functions on the torus too.
These results can be checked with Monte-Carlo simulations, and it would be interesting if one could compute those quantities numerically and verify our results.
However be warned that the result can only be checked by running a simulation at quite a low temperature, due to the presence of the soft modes of \cite{Alvarez-Gaume:2016vff}.

The appearance of perturbatively calculable disorder is interesting, and should make it possible to be far more
explicit about the long-distance behavior of the $O(4)$ model (and other $O(2N)$ models) at finite chemical potential. 
In principle one could calculate explicitly which sets of operators obey cluster decomposition and which do not.  However 
such a calculation would take a more thorough study of the classical solutions than we have performed so far.
For instance, one would have to be more careful about the possibility of ground states preserving nontrivial combinations
of translational and internal global symmetries while breaking each separately; we have ignored those possible breaking
patterns in this paper.

In future work we hope to study the large-charge
EFT on the sphere, in order to compute ground-state conformal dimensions via radial quantization.\ok
There the symmetry braking pattern will be much more interesting because of the non-Abelian nature of the symmetry group on the sphere spatial slice.
Also, the accuracy of the large-$J$ expansion
can be improved by computing subleading corrections from higher-derivative operators in the EFT and quantum effects.
\ok

\section*{Acknowledgements}
\begin{sloppypar}
The authors are grateful to Domenico Orlando and Susanne Reffert for valuable discussions.  
SM and MW acknowledge the support by JSPS Research Fellowship for Young Scientists.  The work of SH is supported by the World Premier
International Research Center Initiative (WPI Initiative), MEXT, Japan; by the JSPS Program for Advancing Strategic
International Networks to Accelerate the Circulation of Talented
Researchers;
and also
supported in part by JSPS KAKENHI Grant Numbers JP22740153, JP26400242. SH
is also grateful to the Physics Department at Harvard University, the Walter Burke Institute for Theoretical Physics at Caltech, and the Galileo Galilei Institute
for generous hospitality while this work was in progress.
\end{sloppypar}


\begin{thebibliography}{99}

\bibitem{Hellerman:2015nra} 
  S.~Hellerman, D.~Orlando, S.~Reffert and M.~Watanabe,
  ``On the CFT Operator Spectrum at Large Global Charge,''
  JHEP {\bf 1512}, 071 (2015)
  doi:10.1007/JHEP12(2015)071
  [arXiv:1505.01537 [hep-th]].


\bibitem{Alvarez-Gaume:2016vff} 
  L.~Alvarez-Gaume, O.~Loukas, D.~Orlando and S.~Reffert,
  ``Compensating strong coupling with large charge,''
  JHEP {\bf 1704}, 059 (2017)
  doi:10.1007/JHEP04(2017)059
  [arXiv:1610.04495 [hep-th]].


\bibitem{Monin:2016jmo} 
  A.~Monin, D.~Pirtskhalava, R.~Rattazzi and F.~K.~Seibold,
  ``Semiclassics, Goldstone Bosons and CFT data,''
  JHEP {\bf 1706}, 011 (2017)
  doi:10.1007/JHEP06(2017)011
  [arXiv:1611.02912 [hep-th]].


\bibitem{Loukas:2016ckj} 
  O.~Loukas,
  ``Abelian scalar theory at large global charge,''
  Fortsch.\ Phys.\  {\bf 65}, no. 9, 1700028 (2017)
  doi:10.1002/prop.201700028
  [arXiv:1612.08985 [hep-th]].


\bibitem{Alday:2016njk} 
  L.~F.~Alday,
  ``Large Spin Perturbation Theory for Conformal Field Theories,''
  Phys.\ Rev.\ Lett.\  {\bf 119}, no. 11, 111601 (2017)
  doi:10.1103/PhysRevLett.119.111601
  [arXiv:1611.01500 [hep-th]].


\bibitem{Alday:2016jfr} 
  L.~F.~Alday,
  ``Solving CFTs with Weakly Broken Higher Spin Symmetry,''
  JHEP {\bf 1710}, 161 (2017)
  doi:10.1007/JHEP10(2017)161
  [arXiv:1612.00696 [hep-th]].


\bibitem{Komargodski:2012ek} 
  Z.~Komargodski and A.~Zhiboedov,
  ``Convexity and Liberation at Large Spin,''
  JHEP {\bf 1311}, 140 (2013)
  doi:10.1007/JHEP11(2013)140
  [arXiv:1212.4103 [hep-th]].


\bibitem{Fitzpatrick:2012yx} 
  A.~L.~Fitzpatrick, J.~Kaplan, D.~Poland and D.~Simmons-Duffin,
  ``The Analytic Bootstrap and AdS Superhorizon Locality,''
  JHEP {\bf 1312}, 004 (2013)
  doi:10.1007/JHEP12(2013)004
  [arXiv:1212.3616 [hep-th]].


\bibitem{Kos:2016ysd} 
  F.~Kos, D.~Poland, D.~Simmons-Duffin and A.~Vichi,
  ``Precision Islands in the Ising and $O(N)$ Models,''
  JHEP {\bf 1608}, 036 (2016)
  doi:10.1007/JHEP08(2016)036
  [arXiv:1603.04436 [hep-th]].


\bibitem{Kos:2015mba} 
  F.~Kos, D.~Poland, D.~Simmons-Duffin and A.~Vichi,
  ``Bootstrapping the O(N) Archipelago,''
  JHEP {\bf 1511}, 106 (2015)
  doi:10.1007/JHEP11(2015)106
  [arXiv:1504.07997 [hep-th]].


\bibitem{Kos:2013tga} 
  F.~Kos, D.~Poland and D.~Simmons-Duffin,
  ``Bootstrapping the $O(N)$ vector models,''
  JHEP {\bf 1406}, 091 (2014)
  doi:10.1007/JHEP06(2014)091
  [arXiv:1307.6856 [hep-th]].


\bibitem{Dey:2016zbg} 
  P.~Dey, A.~Kaviraj and K.~Sen,
  ``More on analytic bootstrap for O(N) models,''
  JHEP {\bf 1606}, 136 (2016)
  doi:10.1007/JHEP06(2016)136
  [arXiv:1602.04928 [hep-th]].


\bibitem{Diab:2016spb} 
  K.~Diab, L.~Fei, S.~Giombi, I.~R.~Klebanov and G.~Tarnopolsky,
  ``On ${C}_{J}$ and ${C}_{T}$ in the Gross–Neveu and O(N) models,''
  J.\ Phys.\ A {\bf 49}, no. 40, 405402 (2016)
  doi:10.1088/1751-8113/49/40/405402
  [arXiv:1601.07198 [hep-th]].


\bibitem{Rattazzi:2008pe} 
  R.~Rattazzi, V.~S.~Rychkov, E.~Tonni and A.~Vichi,
  ``Bounding scalar operator dimensions in 4D CFT,''
  JHEP {\bf 0812}, 031 (2008)
  doi:10.1088/1126-6708/2008/12/031
  [arXiv:0807.0004 [hep-th]].


\bibitem{ElShowk:2012ht} 
  S.~El-Showk, M.~F.~Paulos, D.~Poland, S.~Rychkov, D.~Simmons-Duffin and A.~Vichi,
  ``Solving the 3D Ising Model with the Conformal Bootstrap,''
  Phys.\ Rev.\ D {\bf 86}, 025022 (2012)
  doi:10.1103/PhysRevD.86.025022
  [arXiv:1203.6064 [hep-th]].


\bibitem{El-Showk:2014dwa} 
  S.~El-Showk, M.~F.~Paulos, D.~Poland, S.~Rychkov, D.~Simmons-Duffin and A.~Vichi,
  ``Solving the 3d Ising Model with the Conformal Bootstrap II. c-Minimization and Precise Critical Exponents,''
  J.\ Stat.\ Phys.\  {\bf 157}, 869 (2014)
  doi:10.1007/s10955-014-1042-7
  [arXiv:1403.4545 [hep-th]].


\bibitem{Hellerman:2017veg} 
  S.~Hellerman, S.~Maeda and M.~Watanabe,
  ``Operator Dimensions from Moduli,''
  JHEP {\bf 1710}, 089 (2017)
  doi:10.1007/JHEP10(2017)089
  [arXiv:1706.05743 [hep-th]].


\bibitem{Hellerman:2017sur} 
  S.~Hellerman and S.~Maeda,
  ``On the Large $R$-charge Expansion in ${\mathcal N} = 2$ Superconformal Field Theories,''
  JHEP {\bf 1712}, 135 (2017)
  doi:10.1007/JHEP12(2017)135
  [arXiv:1710.07336 [hep-th]].


\bibitem{Hellerman:2018xpi} 
  S.~Hellerman, S.~Maeda, D.~Orlando, S.~Reffert and M.~Watanabe,
  ``Universal correlation functions in rank 1 SCFTs,''
  arXiv:1804.01535 [hep-th].


\bibitem{Hellerman:2017efx} 
  S.~Hellerman, N.~Kobayashi, S.~Maeda and M.~Watanabe,
  ``A Note on Inhomogeneous Ground States at Large Global Charge,''
  arXiv:1705.05825 [hep-th].


\bibitem{Cuomo:2017vzg} 
  G.~Cuomo, A.~de la Fuente, A.~Monin, D.~Pirtskhalava and R.~Rattazzi,
  ``Rotating superfluids and spinning charged operators in conformal field theory,''
  Phys.\ Rev.\ D {\bf 97}, no. 4, 045012 (2018)
  doi:10.1103/PhysRevD.97.045012
  [arXiv:1711.02108 [hep-th]].


\bibitem{Gross:1987ar} 
  D.~J.~Gross and P.~F.~Mende,
  ``String Theory Beyond the Planck Scale,''
  Nucl.\ Phys.\ B {\bf 303}, 407 (1988).
  doi:10.1016/0550-3213(88)90390-2


\bibitem{Gross:1987kza} 
  D.~J.~Gross and P.~F.~Mende,
  ``The High-Energy Behavior of String Scattering Amplitudes,''
  Phys.\ Lett.\ B {\bf 197}, 129 (1987).
  doi:10.1016/0370-2693(87)90355-8


\bibitem{Polchinski:1991ax} 
  J.~Polchinski and A.~Strominger,
  ``Effective string theory,''
  Phys.\ Rev.\ Lett.\  {\bf 67}, 1681 (1991).
  doi:10.1103/PhysRevLett.67.1681


\bibitem{Baker:1999xn} 
  M.~Baker and R.~Steinke,
  ``An Effective string theory of Abrikosov-Nielsen-Olesen vortices,''
  Phys.\ Lett.\ B {\bf 474}, 67 (2000)
  doi:10.1016/S0370-2693(00)00007-1
  [hep-ph/9905375].


\bibitem{Sonnenschein:2018aqf} 
  J.~Sonnenschein and D.~Weissman,
  ``Quantizing the rotating string with massive endpoints,''
  arXiv:1801.00798 [hep-th].


\bibitem{Hellerman:2017upi} 
  S.~Hellerman and S.~Maeda,
  ``On Vertex Operators in Effective String Theory,''
  arXiv:1701.06406 [hep-th].


\bibitem{Hellerman:2014cba} 
  S.~Hellerman, S.~Maeda, J.~Maltz and I.~Swanson,
  ``Effective String Theory Simplified,''
  JHEP {\bf 1409}, 183 (2014)
  doi:10.1007/JHEP09(2014)183
  [arXiv:1405.6197 [hep-th]].
  
\bibitem{Hellerman:2013kba} 
  S.~Hellerman and I.~Swanson,
  Phys.\ Rev.\ Lett.\  {\bf 114}, no. 11, 111601 (2015)
  doi:10.1103/PhysRevLett.114.111601
  [arXiv:1312.0999 [hep-th]].


\bibitem{Wilson:1971dc} 
  K.~G.~Wilson and M.~E.~Fisher,
  ``Critical exponents in 3.99 dimensions,''
  Phys.\ Rev.\ Lett.\  {\bf 28}, 240 (1972).
  doi:10.1103/PhysRevLett.28.240


\bibitem{Banerjee:2017fcx} 
  D.~Banerjee, S.~Chandrasekharan and D.~Orlando,
  ``Conformal dimensions via large charge expansion,''
  Phys.\ Rev.\ Lett.\  {\bf 120}, no. 6, 061603 (2018)
  doi:10.1103/PhysRevLett.120.061603
  [arXiv:1707.00711 [hep-lat]].


\bibitem{Watanabe:2013uya} 
  H.~Watanabe, T.~Brauner and H.~Murayama,
  ``Massive Nambu-Goldstone Bosons,''
  Phys.\ Rev.\ Lett.\  {\bf 111}, no. 2, 021601 (2013)
  doi:10.1103/PhysRevLett.111.021601
  [arXiv:1303.1527 [hep-th]].


\bibitem{Hellerman:2006zs} 
  S.~Hellerman, A.~Henriques, T.~Pantev, E.~Sharpe and M.~Ando,
  ``Cluster decomposition, T-duality, and gerby CFT's,''
  Adv.\ Theor.\ Math.\ Phys.\  {\bf 11}, no. 5, 751 (2007)
  doi:10.4310/ATMP.2007.v11.n5.a2
  [hep-th/0606034].


\bibitem{Caldararu:2007tc} 
  A.~Caldararu, J.~Distler, S.~Hellerman, T.~Pantev and E.~Sharpe,
  ``Non-birational twisted derived equivalences in abelian GLSMs,''
  Commun.\ Math.\ Phys.\  {\bf 294}, 605 (2010)
  doi:10.1007/s00220-009-0974-2
  [arXiv:0709.3855 [hep-th]].


\bibitem{Hellerman:2010fv} 
  S.~Hellerman and E.~Sharpe,
  ``Sums over topological sectors and quantization of Fayet-Iliopoulos parameters,''
  Adv.\ Theor.\ Math.\ Phys.\  {\bf 15}, 1141 (2011)
  doi:10.4310/ATMP.2011.v15.n4.a7
  [arXiv:1012.5999 [hep-th]].












\end{thebibliography}

\end{document}